# Dispersion Operators Algebra and Linear Canonical Transformations

**Raoelina Andriambololona[1], Ravo Tokiniaina Ranaivoson[2], Randriamisy Hasimbola Damo Emile[3], Hanitriarivo Rakotoson[4]**

*raoelinasp@yahoo.fr[1];jacquelineraoelina@hotmail.com[1];raoelina.andriambololona@gmail.com[1]; tokhiniaina@gmail.com[2], hasinadamo@gmail.com[3],infotsara@gmail.com[4]*

*Theoretical Physics Department*

*Institut National des Sciences et Techniques Nucléaires (INSTN- Madagascar)*

BP 4279 101-Antananarivo –MADAGASCAR, *instn@moov.mg*

***Abstract***: This work intends to present a study on relations between a Lie algebra called dispersion operators algebra, linear canonical transformation and a phase space representation of quantum mechanics that we have introduced and studied in previous works. The paper begins with a brief recall of our previous works followed by the description of the dispersion operators algebra which is performed in the framework of the phase space representation. Then, linear canonical transformations are introduced and linked with this algebra. A multidimensional generalization of the obtained results is given.

## 1 Introduction

The present work can be considered as a part of a series of studies concerning phase space, linear canonical transformation and quantum theory that we have started and performed in our previous works [1],[2]. Through history and scientific literature, it can be remarked that the description of phase space in quantum theory and related problems like study of canonical transformations are among of the most interesting subjects. We may quote many works since the beginning of quantum physics and until nowadays; for instance we have [3-14]. A well known approach to tackle these problems is based on the utilization of the Wigner distribution but other approaches may be also considered. Our work is in this framework.

Through all the paper, we use the natural system of units in which the light speed $c$ and the reduced Planck constant $\hbar$ are set to unity ($c = 1$ , $\hbar = 1$). We use also bold faced letter to denote operators and normal letter for the eigenvalues. The matricial and tensorial notations used in the section 4 are those of the reference [15].

The main result that we have obtained from [1] was the establishment of a phase space representation of quantum mechanics which takes into account the uncertainty relation. It is based on the introduction of quantum states, denoted $|n, X, P, \Delta p\rangle$, defined by the means values $X, P$ and statistical dispersions $(\Delta x_n)^2 = (2n+1)(\Delta x)^2, (\Delta p_n)^2 = (2n+1)(\Delta p)^2$ of coordinate $x$ and momentum $p$. $\Delta x$ and $\Delta p$ satisfying the relation $(\Delta x)(\Delta p) = \frac{1}{2}$.

For the sake of simplicity of writing, we will use the notation $a = \Delta x$, $\mathscr{b} = \Delta p, \mathcal{A} = (a)^2 = (\Delta x)^2$, and $\mathcal{B} = (\mathscr{b})^2 = (\Delta p)^2$. For instance, the state $|n, X, P, \Delta p\rangle$ will be denoted by $|n, X, P, \mathscr{b}\rangle$. The wave functions corresponding to a state $|n, X, P, \mathscr{b}\rangle = |n, X, P, \Delta p\rangle$ respectively in coordinate and momentum representation are the Harmonic Hermite-Gaussian functions denoted by

$$\langle x|n,X,P,\mathscr{b}\rangle = \varphi_n(x,X,P,\mathscr{b}) = \varphi_n(x,X,P,\Delta p)$$

and their Fourier transform denoted by

$$\langle p|n,X,P,\mathscr{b}\rangle = \tilde{\varphi}_n(p,X,P,\mathscr{b}) = \tilde{\varphi}_n(p,X,P,\Delta p).$$

These functions were introduced and used in our previous works [1], [2], [16]. Explicitly we have

$$\langle x|n,X,P,\mathscr{b}\rangle = \varphi_n(x,X,P,\mathscr{b}) = \frac{H_n(\frac{x-X}{\sqrt{2\mathcal{A}}})}{\sqrt{2^n n!\sqrt{2\pi\mathcal{A}}}} e^{-\frac{(x-X)^2}{4\mathcal{A}}+iPx} \tag{1}$$

$$\langle p|n,X,P,\mathscr{b}\rangle = \tilde{\varphi}_n(p,X,P,\mathscr{b}) = \frac{H_n(\frac{p-P}{\sqrt{2\mathcal{B}}})}{\sqrt{2^n n!\sqrt{2\pi\mathcal{B}}}} e^{-\frac{(p-P)^2}{4\mathcal{B}}-iX(p-P)} \tag{2}$$

As usual $H_n$ is the Hermite polynomial of degree $n$. $\Delta x = \mathscr{a}$ , $\Delta p = \mathscr{b}$, $\mathcal{A} = (\Delta x)^2 = (\mathscr{a})^2 =$ and $\mathcal{B} = (\Delta p)^2 = (\mathscr{b})^2$ satisfying the relations

$$(\Delta x)(\Delta p) = \mathscr{a}\mathscr{b} = \frac{1}{2} \tag{3}$$

$$(\Delta x)^2(\Delta p)^2 = (\mathscr{a})^2(\mathscr{b})^2 = \mathcal{A}\mathcal{B} = \frac{1}{4} \tag{4}$$

The state $|n,X,P,\mathscr{b}\rangle$ is the eigenstate of the coordinate and momentum dispersion operators $\boldsymbol{\Sigma}_{\boldsymbol{p}}$ and $\boldsymbol{\Sigma}_{\boldsymbol{x}}$ respectively with the eigenvalues $(2n+1)(\mathscr{b})^2$ and $(2n+1)(\mathscr{a})^2$ **.** If we denote respectively $\boldsymbol{p}$ and $\boldsymbol{x}$ the momentum and coordinate operators [1], [2], [17] we have [1]

$$\boldsymbol{\Sigma}_{\boldsymbol{p}} = \frac{1}{2}\left[\frac{(\boldsymbol{p}-P)^2}{(\mathscr{b})^2} + \frac{(\boldsymbol{x}-X)^2}{(\mathscr{a})^2}\right](\mathscr{b})^2 = \frac{1}{2}[(\boldsymbol{p}-P)^2 + 4(\mathcal{B})^2(\boldsymbol{x}-X)^2] \tag{5}$$

$$\boldsymbol{\Sigma}_{\boldsymbol{x}} = \frac{1}{2}\left[\frac{(\boldsymbol{p}-P)^2}{(\mathscr{b})^2} + \frac{(\boldsymbol{x}-X)^2}{(\mathscr{a})^2}\right](\mathscr{a})^2 = \frac{1}{2}[4(\mathcal{A})^2(\boldsymbol{p}-P)^2 + (\boldsymbol{x}-X)^2] \tag{6}$$

$$\boldsymbol{\Sigma}_{\boldsymbol{p}}|n,X,P,\mathscr{b}\rangle = (2n+1)(\mathscr{b})^2|n,X,P,\mathscr{b}\rangle = (2n+1)\mathcal{B}|n,X,P,\mathscr{b}\rangle \tag{7}$$

$$\boldsymbol{\Sigma}_{\boldsymbol{x}}|n,X,P,\mathscr{b}\rangle = (2n+1)(\mathscr{a})^2|n,X,P,\mathscr{b}\rangle = (2n+1)\mathcal{A}|n,X,P,\mathscr{b}\rangle \tag{8}$$

In our work [2], it was remarked that a link may be established between linear canonical transformation and the phase space representation of quantum mechanics. In the present work, we show that this link can be described properly with the introduction of a Lie algebra that we may call dispersion operators algebra. This Lie algebra is generated by the dispersion operators and some other operators related to them. We have remarked during the design of the present work that operators analogous to these operators have been already introduced and studied previously in various works on linear canonical transformation [18]. As mentioned, our main contribution in this paper is the exploitation of some properties of these operators in the introduction of the dispersion operators algebra to describe properly the relations that can be established between this algebra, the phase space representation and linear canonical transformation. We introduce also a generalization of the results for the case of multidimensional theory.

## 2 Dispersion Operators Algebra

### 2.1 Definitions and properties

Let us consider the three hermitian operators

$$\begin{cases} \beth^{+} = \boldsymbol{\Sigma}_{\boldsymbol{p}} = \frac{1}{2}[(\boldsymbol{p}-P)^2 + 4(\mathcal{B})^2(\boldsymbol{x}-X)^2] \\ \beth^{-} = \frac{1}{2}[(\boldsymbol{p}-P)^2 - 4(\mathcal{B})^2(\boldsymbol{x}-X)^2] \\ \beth^{\times} = \mathcal{B}[(\boldsymbol{p}-P)(\boldsymbol{x}-X) + (\boldsymbol{x}-X)(\boldsymbol{p}-P)] \end{cases} \tag{9}$$

Using the commutation relation of $\boldsymbol{x}$ and ,$[\boldsymbol{x},\boldsymbol{p}]_- = i$, we can deduce the following commutation relations

$$\begin{cases} [\beth^+, \beth^-]_- = 4i\mathcal{B}\beth^\times \\ [\beth^-, \beth^\times]_- = -4i\mathcal{B}\beth^+ \\ [\beth^\times, \beth^+]_- = 4i\mathcal{B}\beth^- \end{cases} \tag{10}$$

Let $\mathfrak{g}$ be the complex vectorial space generated by the linear combination

$$\beth = \lambda\beth^+ + \mu\beth^- + \nu\beth^\times$$

of the three operators $\beth^+, \beth^-$ and $\beth^\times$ with $\lambda, \mu, \nu$ three complex numbers. It can be deduced easily form the relation (10) that $\mathfrak{g}$ is a complex Lie algebra of three dimensions. For any two elements

$$\beth_1 = \lambda_1\beth^+ + \mu_1\beth^- + \nu_1\beth^\times$$

$$\beth_2 = \lambda_2\beth^+ + \mu_2\beth^- + \nu_2\beth^\times$$

of $\mathfrak{g}$, it may be shown that

$$\beth_3 = [\beth_1, \beth_2]_- = \lambda_3\beth^+ + \mu_3\beth^- + \nu_3\beth^\times$$

with

$$\begin{cases} \lambda_3 = -4i\mathcal{B}(\mu_1\nu_2 - \nu_1\mu_2) \\ \mu_3 = 4i\mathcal{B}(\nu_1\lambda_2 - \nu_2\lambda_2) \\ \nu_3 = 4i\mathcal{B}(\lambda_1\mu_2 - \lambda_2\mu_1) \end{cases}$$

is also an element of $\mathfrak{g}$. The Lie algebra $\mathfrak{g}$ may be called dispersion operators algebra.

For future use and convenience, we introduce the following operators

$$\begin{cases} \boldsymbol{z}^- = \frac{1}{\sqrt{2}}[(\boldsymbol{p} - P) - 2i\mathcal{B}(\boldsymbol{x} - X)] \\ \boldsymbol{z}^+ = \frac{1}{\sqrt{2}}[(\boldsymbol{p} - P) + 2i\mathcal{B}(\boldsymbol{x} - X)] = (\boldsymbol{z}^-)^\dagger \end{cases} \Leftrightarrow \begin{cases} (\boldsymbol{p} - P) = \frac{1}{\sqrt{2}}(\boldsymbol{z}^- + \boldsymbol{z}^+) \\ (\boldsymbol{x} - X) = \frac{i}{2\sqrt{2}\mathcal{B}}(\boldsymbol{z}^- - \boldsymbol{z}^+) \end{cases} \tag{11}$$

$$\begin{cases} \not{\boldsymbol{x}} = \frac{(\boldsymbol{x} - X)}{\sqrt{2}(\Delta x)} = \frac{(\boldsymbol{x} - X)}{\sqrt{2}\mathscr{a}} = \frac{(\boldsymbol{x} - X)}{\sqrt{2}\mathcal{A}} = \sqrt{2}(\Delta p)(\boldsymbol{x} - X) = \sqrt{2}\mathscr{b}(\boldsymbol{x} - X) = \sqrt{2}\mathcal{B}(\boldsymbol{x} - X) \\ \not{\boldsymbol{p}} = \frac{(\boldsymbol{p} - P)}{\sqrt{2}(\Delta p)} = \frac{(\boldsymbol{p} - P)}{\sqrt{2}\mathscr{b}} = \frac{(\boldsymbol{p} - P)}{\sqrt{2}\mathcal{B}} = \sqrt{2}(\Delta x)(\boldsymbol{p} - P) = \sqrt{2}\mathscr{a}(\boldsymbol{p} - P) = \sqrt{2}\mathcal{A}(\boldsymbol{p} - P) \end{cases} \tag{12}$$

$$\begin{cases} \not{\boldsymbol{z}}^- = \frac{\boldsymbol{z}^-}{\sqrt{2\mathcal{B}}} = \frac{1}{\sqrt{2}}\left[\frac{(\boldsymbol{p} - P)}{\sqrt{2\mathcal{B}}} - i\frac{(\boldsymbol{x} - X)}{\sqrt{2\mathcal{A}}}\right] = \frac{1}{\sqrt{2}}[\not{\boldsymbol{p}} - i\not{\boldsymbol{x}}] \\ \not{\boldsymbol{z}}^+ = \frac{\boldsymbol{z}^+}{\sqrt{2\mathcal{B}}} = \frac{1}{\sqrt{2}}\left[\frac{(\boldsymbol{p} - P)}{\sqrt{2\mathcal{B}}} + i\frac{(\boldsymbol{x} - X)}{\sqrt{2\mathcal{A}}}\right] = \frac{1}{\sqrt{2}}[\not{\boldsymbol{p}} - i\not{\boldsymbol{x}}] \end{cases} \tag{13}$$

The following relations can be deduced

$$\begin{cases} \beth^+ = \mathcal{B}((\not{\boldsymbol{p}})^2 + (\not{\boldsymbol{x}})^2) = \mathcal{B}(\not{\boldsymbol{z}}^-\not{\boldsymbol{z}}^+ + \not{\boldsymbol{z}}^+\not{\boldsymbol{z}}^-) \\ \beth^- = \mathcal{B}((\not{\boldsymbol{p}})^2 - (\not{\boldsymbol{x}})^2) = \mathcal{B}((\not{\boldsymbol{z}}^-)^2 + (\not{\boldsymbol{z}}^+)^2) \\ \beth^\times = \mathcal{B}(\not{\boldsymbol{p}}\not{\boldsymbol{x}} + \not{\boldsymbol{x}}\not{\boldsymbol{p}}) = i\mathcal{B}((\not{\boldsymbol{z}}^-)^2 - (\not{\boldsymbol{z}}^+)^2) \end{cases} \tag{14}$$

We may remark that the sets

$$\{(\boldsymbol{p})^2, (\boldsymbol{x})^2, \boldsymbol{px} + \boldsymbol{xp}\}, \{(\not{\boldsymbol{p}})^2, (\not{\boldsymbol{x}})^2, \not{\boldsymbol{p}}\not{\boldsymbol{x}} + \not{\boldsymbol{x}}\not{\boldsymbol{p}}\}$$

$$\{\boldsymbol{z}^-\boldsymbol{z}^+ + \boldsymbol{z}^+\boldsymbol{z}^-, (\boldsymbol{z}^-)^2, (\boldsymbol{z}^+)^2\}, \{\not{\boldsymbol{z}}^-\not{\boldsymbol{z}}^+ + \not{\boldsymbol{z}}^+\not{\boldsymbol{z}}^-, (\not{\boldsymbol{z}}^-)^2, (\not{\boldsymbol{z}}^+)^2\}$$

are also four basis of the dispersion operators algebra $\mathfrak{g}$.

We have the following commutation relations

$$[\boldsymbol{z}^-, \boldsymbol{z}^+]_- = 2\mathcal{B} \tag{15}$$

$$[\not{\boldsymbol{x}}, \not{\boldsymbol{p}}]_- = \left[\frac{(\boldsymbol{x} - X)}{\sqrt{2}\mathscr{a}}, \frac{(\boldsymbol{p} - P)}{\sqrt{2}\mathscr{b}}\right]_- = [(\boldsymbol{x} - X), (\boldsymbol{p} - P)]_- = [\boldsymbol{x}, \boldsymbol{p}]_- = i \tag{16}$$

$$[\not{\boldsymbol{z}}^-, \not{\boldsymbol{z}}^+] = \frac{1}{2\mathcal{B}}[\boldsymbol{z}^-, \boldsymbol{z}^+] = 1 \tag{17}$$

$$\begin{cases} [(\boldsymbol{x})^2, \boldsymbol{p}\,]_- = 2i\boldsymbol{x} \\ [\boldsymbol{px}, \boldsymbol{p}\,]_- = i\boldsymbol{p} \\ [\boldsymbol{xp}, \boldsymbol{p}\,]_- = i\boldsymbol{p} \end{cases} \tag{18}$$

$$\begin{cases} [(\boldsymbol{p})^2, \boldsymbol{x}\,]_- = -2i\boldsymbol{p} \\ [\boldsymbol{px}, \boldsymbol{x}\,]_- = -i\boldsymbol{x} \\ [\boldsymbol{xp}, \boldsymbol{x}\,]_- = -i\boldsymbol{x} \end{cases} \tag{19}$$

$$\begin{cases} [(\boldsymbol{p})^2, (\boldsymbol{x})^2\,]_- = -2i(\boldsymbol{px} + \boldsymbol{xp}) \\ [(\boldsymbol{p})^2, \boldsymbol{px}\,]_- = -2i(\boldsymbol{p})^2 \\ [(\boldsymbol{x})^2\,, \boldsymbol{px}\,]_- = 2i(\boldsymbol{x})^2 \end{cases} \tag{20}$$

We may define the operators

$$\begin{cases} \bar{\beth}^+ = \dfrac{\beth^+}{4\mathcal{B}} = \dfrac{1}{4}((\boldsymbol{p})^2 + (\boldsymbol{x})^2) = \dfrac{1}{4}(\boldsymbol{z}^-\boldsymbol{z}^+ + \boldsymbol{z}^+\boldsymbol{z}^-) \\ \bar{\beth}^- = \dfrac{\beth^-}{4\mathcal{B}} = \dfrac{1}{4}((\boldsymbol{p})^2 + (\boldsymbol{x})^2) = \dfrac{1}{4}((\boldsymbol{z}^-)^2 + (\boldsymbol{z}^+)^2) \\ \bar{\beth}^\times = \dfrac{\beth^\times}{4\mathcal{B}} = \dfrac{1}{4}(\boldsymbol{px} + \boldsymbol{xp}) = \dfrac{i}{4}((\boldsymbol{z}^-)^2 - (\boldsymbol{z}^+)^2) \end{cases} \tag{21}$$

The set $\{\bar{\beth}^+, \bar{\beth}^-, \bar{\beth}^\times\}$ is also a basis of the dispersion operator algebra $\mathfrak{g}$.It may be deduced easily from the commutation relation of $\beth^+, \beth^-$and $\beth^\times$that $\bar{\beth}^+, \bar{\beth}^-$and $\bar{\beth}^\times$satisfy to the following commutation relations

$$\begin{cases} [\bar{\beth}^+, \bar{\beth}^-]_- = i\bar{\beth}^\times \\ [\bar{\beth}^-, \bar{\beth}^\times]_- = -i\bar{\beth}^+ \\ [\bar{\beth}^\times, \bar{\beth}^+]_- = i\bar{\beth}^- \end{cases} \tag{22}$$

And we have

$$\begin{cases} [\bar{\beth}^+, \boldsymbol{p}]_- = \dfrac{1}{2}i\boldsymbol{x} \\ [\bar{\beth}^-, \boldsymbol{p}]_- = -\dfrac{1}{2}i\boldsymbol{x} \\ [\bar{\beth}^\times, \boldsymbol{p}]_- = \dfrac{1}{2}i\boldsymbol{p} \end{cases} \tag{23}$$

$$\begin{cases} [\bar{\beth}^+, \boldsymbol{x}]_- = -\dfrac{1}{2}i\boldsymbol{p} \\ [\bar{\beth}^-, \boldsymbol{x}]_- = -\dfrac{1}{2}i\boldsymbol{p} \\ [\bar{\beth}^\times, \boldsymbol{x}]_- = -\dfrac{1}{2}i\boldsymbol{x} \end{cases} \tag{24}$$

### 2.2 Representation of the dispersion operators algebra over the state space of a particle

To define a representation of the dispersion operators algebra $\mathfrak{g}$ over the state space $\mathcal{E}$ of a particle, we have to find the matrix representation of the three operators $\beth^+, \beth^-$and $\beth^\times$ in a basis of $\mathcal{E}$. It is obvious that an adequate basis is the basis $\{|n, X, P, \mathscr{b}\rangle\}$ composed by the eigenstates of the momentum dispersion operator $\beth^+$

$$\beth^+|n, X, P, \mathscr{b}\rangle = (2n+1)\mathcal{B}|n, X, P, \mathscr{b}\rangle \tag{25}$$

$\beth^+$is represented by a diagonal matrix with elements equal to $(2n+1)\mathcal{B}$.

Now we have to find the expressions of $\beth^-|n, X, P, \mathscr{b}\rangle$ and $\beth^\times|n, X, P, \mathscr{b}\rangle$. We consider the relation

$$\begin{cases} \beth^+ = \mathcal{B}(\boldsymbol{z}^-\boldsymbol{z}^+ + \boldsymbol{z}^+\boldsymbol{z}^-) \\ \beth^- = \mathcal{B}[(\boldsymbol{z}^-)^2 + (\boldsymbol{z}^+)^2] \\ \beth^\times = i\mathcal{B}[(\boldsymbol{z}^-)^2 - (\boldsymbol{z}^+)^2] \end{cases} \tag{26}$$

Let us first search for the expression of $\boldsymbol{z}^-|n, X, P, \mathscr{b}\rangle$ and $\boldsymbol{z}^+|n, X, P, \mathscr{b}\rangle$. From the commutation relation $[\boldsymbol{z}^-, \boldsymbol{z}^+]_- = 1$ we can deduce the relation

$$\beth^+ = \mathcal{B}(\boldsymbol{z}^-\boldsymbol{z}^+ + \boldsymbol{z}^+\boldsymbol{z}^-) = \mathcal{B}(2\boldsymbol{z}^+\boldsymbol{z}^- + 1) = \mathcal{B}(2\boldsymbol{z}^-\boldsymbol{z}^+ - 1) \tag{27}$$

and the commutation relations

$$\begin{cases} [\beth^+, \boldsymbol{\mathscr{z}}^-]_- = -2\mathcal{B}\boldsymbol{\mathscr{z}}^- \\ [\beth^+, \boldsymbol{\mathscr{z}}^+]_- = 2\mathcal{B}\boldsymbol{\mathscr{z}}^+ \end{cases} \tag{28}$$

Then from the relations (27) and (28), it may be deduced after lengthy but straightforward calculations that

$$\begin{cases} \boldsymbol{\mathscr{z}}^- |n, X, P, \mathscr{b}\rangle = \sqrt{n}|n-1, X, P, \mathscr{b}\rangle \\ \boldsymbol{\mathscr{z}}^+ |n, X, P, \mathscr{b}\rangle = \sqrt{n+1}|n+1, X, P, \mathscr{b}\rangle \end{cases} \tag{29}$$

So

$$\begin{cases} (\boldsymbol{\mathscr{z}}^-)^2 |n, X, P, \mathscr{b}\rangle = \boldsymbol{\mathscr{z}}^- \sqrt{n}|n-1, X, P, \mathscr{b}\rangle = \sqrt{n(n-1)}|n-2, X, P, \mathscr{b}\rangle \\ (\boldsymbol{\mathscr{z}}^+)^2 |n, X, P, \mathscr{b}\rangle = \boldsymbol{\mathscr{z}}^+ \sqrt{n+1}|n-1, X, P, \mathscr{b}\rangle = \sqrt{(n+1)(n+2)}|n+2, X, P, \mathscr{b}\rangle \end{cases} \tag{30}$$

As

$$\begin{cases} \beth^- = \mathcal{B}[(\boldsymbol{\mathscr{z}}^-)^2 + (\boldsymbol{\mathscr{z}}^+)^2] \\ \beth^\times = i\mathcal{B}[(\boldsymbol{\mathscr{z}}^-)^2 - (\boldsymbol{\mathscr{z}}^+)^2] \end{cases} \tag{31}$$

we obtain for the representation of the three operators $\beth^+, \beth^-$ and $\beth^\times$ in the basis $\{|n, X, P, \mathscr{b}\rangle\}$ of the state space $\mathcal{E}$ of a particle

$$\begin{cases} \beth^+ |n, X, P, \mathscr{b}\rangle = (2n+1)\mathcal{B}|n, X, P, \mathscr{b}\rangle \\ \beth^- |n, X, P, \mathscr{b}\rangle = \sqrt{n(n-1)}\mathcal{B}|n-2, X, P, \mathscr{b}\rangle + \sqrt{(n+1)(n+2)}\mathcal{B}|n+2, X, P, \mathscr{b}\rangle \\ \beth^\times |n, X, P, \mathscr{b}\rangle = i\left[\sqrt{n(n-1)}\mathcal{B}|n-2, X, P, \mathscr{b}\rangle - \sqrt{(n+1)(n+2)}\mathcal{B}|n+2, X, P, \mathscr{b}\rangle\right] \end{cases} \tag{32}$$

We may also write these relations in the form

$$\begin{cases} \beth^+ |n, X, P, \mathscr{b}\rangle = (2n+1)\mathcal{B}|n, X, P, \mathscr{b}\rangle \\ (\beth^- - i\beth^\times)|n, X, P, \mathscr{b}\rangle = 2\sqrt{n(n-1)}\mathcal{B}|n-2, X, P, \mathscr{b}\rangle \\ (\beth^- + i\beth^\times)|n, X, P, \mathscr{b}\rangle = 2\sqrt{(n+1)(n+2)}\mathcal{B}|n+2, X, P, \mathscr{b}\rangle \end{cases} \tag{33}$$

## 3 Linear Canonical Transformations

### 3.1 Definitions and properties

In quantum mechanics, a linear canonical transformation can be defined as a linear transformation mixing the coordinate operator $\boldsymbol{x}$ and the momentum operator $\boldsymbol{p}$ and leaving invariant the commutator $[\boldsymbol{x}, \boldsymbol{p}] = i$ [2]. As $\boldsymbol{x}$ and $\boldsymbol{p}$ are linked with the operators $\boldsymbol{\mathscr{x}}$ and $\boldsymbol{\mathscr{p}}$ through the linear relations (12), we may also take a definition of linear canonical transformation as linear transformation mixing $\boldsymbol{\mathscr{x}}$ and $\boldsymbol{\mathscr{p}}$

$$\begin{cases} \boldsymbol{\mathscr{p}}' = \Pi\boldsymbol{\mathscr{p}} + \Theta\boldsymbol{\mathscr{x}} \\ \boldsymbol{\mathscr{x}}' = \Xi\boldsymbol{\mathscr{p}} + \Lambda\boldsymbol{\mathscr{x}} \end{cases} \Leftrightarrow (\boldsymbol{\mathscr{p}}' \quad \boldsymbol{\mathscr{x}}') = (\boldsymbol{\mathscr{p}} \quad \boldsymbol{\mathscr{x}})\begin{pmatrix} \Pi & \Xi \\ \Theta & \Lambda \end{pmatrix} \tag{34}$$

in which

$$\begin{cases} \boldsymbol{\mathscr{x}}' = \dfrac{(\boldsymbol{x}' - X')}{\sqrt{2}a'} = \dfrac{(\boldsymbol{x}' - X')}{\sqrt{2\mathcal{A}'}} = \sqrt{2}\mathscr{b}'(\boldsymbol{x}' - X') = \sqrt{2\mathcal{B}'}(\boldsymbol{x}' - X') \\ \boldsymbol{\mathscr{p}}' = \dfrac{(\boldsymbol{p}' - P')}{\sqrt{2}\mathscr{b}'} = \dfrac{(\boldsymbol{p}' - P')}{\sqrt{2\mathcal{B}'}} = \sqrt{2}a'(\boldsymbol{p}' - P') = \sqrt{2\mathcal{A}'}(\boldsymbol{p}' - P') \end{cases} \tag{35}$$

$$[\boldsymbol{\mathscr{x}}', \boldsymbol{\mathscr{p}}']_- = \frac{1}{2a'\mathscr{b}'}[(\boldsymbol{x}' - X'), (\boldsymbol{p}' - P')]_- = [\boldsymbol{x}', \boldsymbol{p}']_- \tag{36}$$

where $\boldsymbol{x}'$ and $\boldsymbol{p}'$ are the new coordinate and momentum operators resulting from the transformation. If we have a linear canonical transformation, we must have

$$[\boldsymbol{x}', \boldsymbol{p}'] = [\boldsymbol{x}, \boldsymbol{p}] = i \tag{37}$$

So taking into account the relation (36), we must have

$$[\boldsymbol{\mathscr{x}}', \boldsymbol{\mathscr{p}}'] = [\boldsymbol{x}', \boldsymbol{p}'] = [\boldsymbol{x}, \boldsymbol{p}] = i = [\boldsymbol{\mathscr{x}}, \boldsymbol{\mathscr{p}}] \tag{38}$$

Then in our case the full definition of the linear canonical transformation is

$$\begin{cases} \boldsymbol{p}' = \Pi\boldsymbol{p} + \Theta\boldsymbol{x} \\ \boldsymbol{x}' = \Xi\boldsymbol{p} + \Lambda\boldsymbol{x} \\ [\boldsymbol{x}', \boldsymbol{p}'] = [\boldsymbol{x}, \boldsymbol{p}] = i \end{cases} \tag{39}$$

The last condition $[\boldsymbol{x}', \boldsymbol{p}'] = [\boldsymbol{x}, \boldsymbol{p}] = i$ leads to the relation

$$\Pi\Lambda - \Theta\Xi = 1 \Leftrightarrow \begin{vmatrix} \Pi & \Xi \\ \Theta & \Lambda \end{vmatrix} = 1 \tag{40}$$

If we consider real linear canonical transformation (the parameters $\Pi$, $\Lambda$, $\Xi$ and $\Theta$ are real) the relation (40) means that the matrix $\begin{pmatrix} \Pi & \Xi \\ \Theta & \Lambda \end{pmatrix}$ is an element of the special linear group $SL(2, \mathbb{R})$. We may write it in the form

$$\begin{pmatrix} \Pi & \Xi \\ \Theta & \Lambda \end{pmatrix} = e^{\mathcal{M}} = e^{\begin{pmatrix} \mathcal{M}_1 & \mathcal{M}_3 \\ \mathcal{M}_2 & \mathcal{M}_4 \end{pmatrix}} \tag{41}$$

with $\mathcal{M}$ an element of the Lie algebra $\mathfrak{sl}(2, \mathbb{R})$ of the Lie group $SL(2, \mathbb{R})$, we have

$$\Pi\Lambda - \Theta\Xi = 1 \Leftrightarrow \mathcal{M}_4 = -\mathcal{M}_1 \Leftrightarrow \mathcal{M} = \begin{pmatrix} \mathcal{M}_1 & \mathcal{M}_3 \\ \mathcal{M}_2 & -\mathcal{M}_1 \end{pmatrix} \tag{42}$$

Then, for an infinitesimal linear canonical transformation, we have

$$\begin{pmatrix} \Pi & \Xi \\ \Theta & \Lambda \end{pmatrix} = 1 + \mathcal{M} = \begin{pmatrix} 1 & 0 \\ 0 & 1 \end{pmatrix} + \begin{pmatrix} \mathcal{M}_1 & \mathcal{M}_3 \\ \mathcal{M}_2 & -\mathcal{M}_1 \end{pmatrix} = \begin{pmatrix} 1 + \mathcal{M}_1 & \mathcal{M}_3 \\ \mathcal{M}_2 & 1 - \mathcal{M}_1 \end{pmatrix} \tag{43}$$

So

$$\begin{cases} \boldsymbol{p}' = \Pi\boldsymbol{p} + \Theta\boldsymbol{x} = \boldsymbol{p} + \mathcal{M}_1\boldsymbol{p} + \mathcal{M}_2\boldsymbol{x} \\ \boldsymbol{x}' = \Xi\boldsymbol{p} + \Lambda\boldsymbol{x} = \boldsymbol{x} + \mathcal{M}_3\boldsymbol{p} - \mathcal{M}_1\boldsymbol{x} \end{cases} \tag{44}$$

### 3.2 Unitary representation and relation with dispersion operators

As the linear canonical transformation is a transformation which affects quantum operators, we may represent it by an unitary transformation

$$\begin{cases} \boldsymbol{p}' = \Pi\boldsymbol{p} + \Theta\boldsymbol{x} = \boldsymbol{U}\boldsymbol{p}\boldsymbol{U}^\dagger \\ \boldsymbol{x}' = \Xi\boldsymbol{p} + \Lambda\boldsymbol{x} = \boldsymbol{U}\boldsymbol{x}\boldsymbol{U}^\dagger \end{cases} \tag{45}$$

where $\boldsymbol{U}$ is a unitary operator which can be considered as acting in the state space $\mathcal{E}$ of a particle. It may be verified that the commutator $[\boldsymbol{x}, \boldsymbol{p}] = i$ is invariant under the unitary representation defined in (45) as expected for a linear canonical transformation

$$[\boldsymbol{x}', \boldsymbol{p}'] = [\boldsymbol{U}\boldsymbol{x}\boldsymbol{U}^\dagger, \boldsymbol{U}\boldsymbol{p}\boldsymbol{U}^\dagger] = \boldsymbol{U}\boldsymbol{x}\boldsymbol{U}^\dagger\boldsymbol{U}\boldsymbol{p}\boldsymbol{U}^\dagger - \boldsymbol{U}\boldsymbol{p}\boldsymbol{U}^\dagger\boldsymbol{U}\boldsymbol{x}\boldsymbol{U}^\dagger = \boldsymbol{U}[\boldsymbol{x}, \boldsymbol{p}]\boldsymbol{U}^\dagger = i\boldsymbol{U}\boldsymbol{U}^\dagger = i \tag{46}$$

It can be shown that $\boldsymbol{U}$ may be written in the form

$$\boldsymbol{U} = e^{i\beth} = e^{i(\theta_+\beth^+ + \theta_-\beth^- + \theta_\times\beth^\times)} \tag{47}$$

in which $\beth = \theta_+\beth^+ + \theta_-\beth^- + \theta_\times\beth^\times$ is an element of the dispersion operators algebra and $\theta_+, \theta_-, \theta_\times$ three real numbers. In fact, for an infinitesimal transformation, we have

$$\boldsymbol{U} = 1 + i(\theta_+\beth^+ + \theta_-\beth^- + \theta_\times\beth^\times) \qquad \boldsymbol{U}^\dagger = 1 - i(\theta_+\beth^+ + \theta_-\beth^- + \theta_\times\beth^\times)$$

Then it can be deduced from (45) that

$$\begin{cases} \boldsymbol{p}' = \boldsymbol{p} + i\theta_+[\beth^+, \boldsymbol{p}]_- + i\theta_-[\beth^-, \boldsymbol{p}]_- + i\theta_\times[\beth^\times, \boldsymbol{p}]_- \\ \boldsymbol{x}' = \boldsymbol{x} + i\theta_+[\beth^+, \boldsymbol{x}]_- + i\theta_-[\beth^-, \boldsymbol{x}]_- + i\theta_\times[\beth^\times, \boldsymbol{x}]_- \end{cases} \tag{48}$$

Taking into account the relations (23) and (24), we obtain

$$\begin{cases} \boldsymbol{p}' = \boldsymbol{p} - \frac{1}{2}\theta_+ \boldsymbol{x} + \frac{1}{2}\theta_- \boldsymbol{x} - \frac{1}{2}\theta_\times \boldsymbol{p} = \boldsymbol{p} - \frac{1}{2}\theta_\times \boldsymbol{p} + \frac{1}{2}(\theta_- - \theta_+)\boldsymbol{x} \\ \boldsymbol{x}' = \boldsymbol{x} + \frac{1}{2}\theta_+ \boldsymbol{p} + \frac{1}{2}\theta_- \boldsymbol{p} + \frac{1}{2}\theta_\times \boldsymbol{x} = \boldsymbol{x} + \frac{1}{2}(\theta_- + \theta_+)\boldsymbol{p} + \frac{1}{2}\theta_\times \boldsymbol{x} \end{cases} \tag{49}$$

identifying the relations (44) and (49) it is deduced

$$\begin{cases} \mathcal{M}_1 = -\frac{1}{2}\theta_\times \\ \mathcal{M}_2 = \frac{1}{2}(\theta_- - \theta_+) \\ \mathcal{M}_3 = \frac{1}{2}(\theta_- + \theta_+) \end{cases} \tag{50}$$

So briefly, we have for a linear canonical transformation

$$\begin{cases} \boldsymbol{p}' = \Pi \boldsymbol{p} + \Theta \boldsymbol{x} = \boldsymbol{U}\boldsymbol{p}\boldsymbol{U}^\dagger \\ \boldsymbol{x}' = \Xi \boldsymbol{p} + \Lambda \boldsymbol{x} = \boldsymbol{U}\boldsymbol{x}\boldsymbol{U}^\dagger \end{cases} \Leftrightarrow (\boldsymbol{p}' \quad \boldsymbol{x}') = (\boldsymbol{p} \quad \boldsymbol{x})\begin{pmatrix} \Pi & \Xi \\ \Theta & \Lambda \end{pmatrix} \tag{51}$$

with

$$\begin{pmatrix} \Pi & \Xi \\ \Theta & \Lambda \end{pmatrix} = e^{\mathcal{M}} = e^{\begin{pmatrix} \mathcal{M}_1 & \mathcal{M}_3 \\ \mathcal{M}_2 & \mathcal{M}_4 \end{pmatrix}} = e^{\frac{1}{2}\begin{pmatrix} -\theta_\times & \theta_+ + \theta_- \\ \theta_- - \theta_+ & \theta_\times \end{pmatrix}} \tag{52}$$

and

$$\boldsymbol{U} = e^{i(\theta_+ \bar{\beth}^+ + \theta_- \bar{\beth}^- + \theta_\times \bar{\beth}^\times)} \tag{53}$$

The unitarity of $\boldsymbol{U}$ results from the hermiticity of the operators $\bar{\beth}^+, \bar{\beth}^-$ and $\bar{\beth}^\times$.

### 3.3 Transformation law of the basis $\{\beth^+, \beth^-, \beth^\times\}$ of the dispersion operators algebra

Taking into account the relation (14) and (21), we may define the operators

$$\begin{cases} \beth^{+\prime} = \mathcal{B}'[(\boldsymbol{p}')^2 + (\boldsymbol{x}')^2] \\ \beth^{-\prime} = \mathcal{B}'[(\boldsymbol{p}')^2 - (\boldsymbol{x}')^2] \\ \beth^{\times\prime} = \mathcal{B}'(\boldsymbol{p}'\boldsymbol{x}' + \boldsymbol{x}'\boldsymbol{p}') \end{cases} \tag{54}$$

$$\begin{cases} \bar{\beth}^{+\prime} = \frac{\beth^{+\prime}}{4\mathcal{B}'} = \frac{1}{4}[(\boldsymbol{p}')^2 + (\boldsymbol{x}')^2] \\ \bar{\beth}^{-\prime} = \frac{\beth^{-\prime}}{4\mathcal{B}'} = \frac{1}{4}[(\boldsymbol{p}')^2 - (\boldsymbol{x}')^2] \\ \bar{\beth}^{\times\prime} = \frac{\beth^{\times\prime}}{4\mathcal{B}'} = \frac{1}{4}(\boldsymbol{p}'\boldsymbol{x}' + \boldsymbol{x}'\boldsymbol{p}') \end{cases} \tag{55}$$

Taking into account the relations (51), we obtain

$$\begin{cases} \bar{\beth}^{+\prime} = \frac{1}{2}[(\Pi)^2 + (\Theta)^2]\bar{\beth}^+ + \frac{1}{2}[(\Xi)^2 - (\Lambda)^2]\bar{\beth}^- + (\Pi\Theta + \Xi\Lambda)\bar{\beth}^\times = \boldsymbol{U}\bar{\beth}^+\boldsymbol{U}^\dagger \\ \bar{\beth}^{-\prime} = \frac{1}{2}[(\Pi)^2 + (\Theta)^2]\bar{\beth}^+ - \frac{1}{2}[(\Xi)^2 - (\Lambda)^2]\bar{\beth}^- + (\Pi\Theta - \Xi\Lambda)\bar{\beth}^\times = \boldsymbol{U}\bar{\beth}^-\boldsymbol{U}^\dagger \\ \bar{\beth}^{\times\prime} = (\Pi\Xi + \Theta\Lambda)\bar{\beth}^+ + (\Pi\Xi - \Theta\Lambda)\bar{\beth}^- + (\Pi\Lambda + \Theta\Xi)\bar{\beth}^\times = \boldsymbol{U}\bar{\beth}^\times\boldsymbol{U}^\dagger \end{cases} \tag{56}$$

$$\begin{cases} \beth^{+\prime} = \frac{1}{2}\frac{\mathcal{B}'}{\mathcal{B}}\{[(\Pi)^2 + (\Theta)^2]\beth^+ + \frac{1}{2}[(\Xi)^2 - (\Lambda)^2]\beth^- + (\Pi\Theta + \Xi\Lambda)\beth^\times\} = \frac{\mathcal{B}'}{\mathcal{B}}\boldsymbol{U}\beth^+\boldsymbol{U}^\dagger \\ \beth^{-\prime} = \frac{1}{2}\frac{\mathcal{B}'}{\mathcal{B}}\{[(\Pi)^2 + (\Theta)^2]\beth^+ - \frac{1}{2}[(\Xi)^2 - (\Lambda)^2]\beth^- + (\Pi\Theta - \Xi\Lambda)\beth^\times\} = \frac{\mathcal{B}'}{\mathcal{B}}\boldsymbol{U}\beth^-\boldsymbol{U}^\dagger \\ \beth^{\times\prime} = \frac{\mathcal{B}'}{\mathcal{B}}\{[(\Pi\Xi + \Theta\Lambda)\beth^+ + (\Pi\Xi - \Theta\Lambda)\beth^- + (\Pi\Lambda + \Theta\Xi)\beth^\times]\} = \frac{\mathcal{B}'}{\mathcal{B}}\boldsymbol{U}\beth^\times\boldsymbol{U}^\dagger \end{cases} \tag{57}$$

## 4 Multidimensional Generalization

### 4.1 Dispersion Operators Algebra

Let $\boldsymbol{p}_\mu$ and $\boldsymbol{x}^\nu$ be respectively the momentum and coordinate components [1]. They obey the commutation relation $[\boldsymbol{p}_\mu, \boldsymbol{x}^\nu]_- = i\delta_\mu^\nu$. We may generalize the operators $\beth^+, \beth^-, \beth^\times$ by the following tensor operators

$$
\begin{cases}
\beth_{\boldsymbol{\mu\nu}}^{+} = \dfrac{1}{2}[(\boldsymbol{p_\mu} - P_\mu)(\boldsymbol{p_\nu} - P_\nu) + 4\mathcal{B}_{\mu\alpha}\mathcal{B}_{\nu\beta}(\boldsymbol{x}^\alpha - X^\alpha)(\boldsymbol{x}^\beta - X^\beta)] \\
\beth_{\boldsymbol{\mu\nu}}^{-} = \dfrac{1}{2}[\ (\boldsymbol{p_\mu} - P_\mu)(\boldsymbol{p_\nu} - P_\nu) - 4\mathcal{B}_{\mu\alpha}\mathcal{B}_{\nu\beta}(\boldsymbol{x}^\alpha - X^\alpha)(\boldsymbol{x}^\beta - X^\beta)] \\
\beth_{\boldsymbol{\mu\nu}}^{\times} = \mathcal{B}_{\mu\alpha}[(\boldsymbol{p_\nu} - P_\nu)(\boldsymbol{x}^\alpha - X^\alpha) + (\boldsymbol{x}^\alpha - X^\alpha)(\boldsymbol{p_\nu} - P_\nu)]
\end{cases}
\quad (58)
$$

in which $\mathcal{B}_{\mu\nu}$ are the components of the momentum dispersion-codispersion tensor [1]. Let $\eta_{\mu\nu}$ be the components of the symmetric bilinear form $\eta$ associated with the considered space. For the case of a general $N$-dimensional pseudo-Euclidian space, if $(N_+, N_-)$ is the signature of $\eta$ $(N_+ + N_- = N)$ , we have

$$
\eta_{\mu\nu} = \begin{cases}
\ \ 1 \quad for\ \mu = \nu = 0,1, \dots, N_+ - 1 \\
-1 \quad for\ \mu = \nu = N_+, N_+ + 1, \dots, N - 1 \\
\ \ 0 \quad for\ \mu \neq \nu
\end{cases}
\quad (59)
$$

for instance, in the case of Minkowski space, the signature of $\eta$ is $(1{,}3)$. So we have

$$
\eta_{\mu\nu} = \begin{cases}
\ \ 1 \quad for\ \mu = \nu = 0 \\
-1 \quad for\ \mu = \nu = 1, 2, 3 \\
\ \ 0 \quad for \mu \neq \nu
\end{cases}
\quad (60)
$$

If we introduce the operators $\boldsymbol{\mathfrak{p}_\mu}$and $\boldsymbol{\mathfrak{x}^\nu}$defined by the relations

$$
\begin{cases}
\boldsymbol{p_\mu} = \sqrt{2}\mathcal{b}_{\boldsymbol{\mu}}^{\boldsymbol{\nu}}\boldsymbol{\mathfrak{p}_\mu} + P_\mu \\
\boldsymbol{x^\nu} = \sqrt{2}\mathcal{a}_{\boldsymbol{\mu}}^{\boldsymbol{\nu}}\boldsymbol{\mathfrak{x}^\mu} + X^\mu
\end{cases}
\quad (61)
$$

with $\mathcal{a}_{\boldsymbol{\mu}}^{\boldsymbol{\nu}}$ and $\mathcal{b}_{\boldsymbol{\mu}}^{\boldsymbol{\nu}}$verifying the relations

$$
\begin{cases}
\mathcal{B}_{\mu\alpha}\mathcal{a}_{\nu}^{\alpha} = \dfrac{1}{2}\mathcal{b}_{\mu\nu} = \eta_{\mu\rho}\mathcal{b}_{\boldsymbol{\nu}}^{\rho} \\
\mathcal{a}_{\boldsymbol{\mu}}^{\lambda}\mathcal{b}_{\lambda}^{\nu} = \dfrac{1}{2}\delta_{\boldsymbol{\mu}}^{\boldsymbol{\nu}}
\end{cases}
\quad (62)
$$

($\delta_{\boldsymbol{\mu}}^{\boldsymbol{\sigma}}$ being the components of the Kronecker's symbol ) we obtain

$$
\begin{cases}
\beth_{\boldsymbol{\mu\nu}}^{+} = \mathcal{b}_{\boldsymbol{\mu}}^{\boldsymbol{\rho}}\mathcal{b}_{\boldsymbol{\nu}}^{\boldsymbol{\lambda}}\boldsymbol{\beth}_{\boldsymbol{\rho\lambda}}^{+} \\
\beth_{\boldsymbol{\mu\nu}}^{-} = \mathcal{b}_{\boldsymbol{\mu}}^{\boldsymbol{\rho}}\mathcal{b}_{\boldsymbol{\nu}}^{\boldsymbol{\lambda}}\boldsymbol{\beth}_{\boldsymbol{\rho\lambda}}^{-} \\
\beth_{\boldsymbol{\mu\nu}}^{\times} = \mathcal{b}_{\boldsymbol{\mu}}^{\boldsymbol{\rho}}\mathcal{b}_{\boldsymbol{\nu}}^{\boldsymbol{\lambda}}\boldsymbol{\beth}_{\boldsymbol{\rho\lambda}}^{\times}
\end{cases}
\quad (63)
$$

in which

$$
\begin{cases}
\boldsymbol{\beth}_{\boldsymbol{\mu\nu}}^{+} = \dfrac{1}{4}(\boldsymbol{\mathfrak{p}}_\mu\boldsymbol{\mathfrak{p}}_\nu + \boldsymbol{\mathfrak{x}}_\mu\boldsymbol{\mathfrak{x}}_\nu) \\
\boldsymbol{\beth}_{\boldsymbol{\mu\nu}}^{-} = \dfrac{1}{4}(\boldsymbol{\mathfrak{p}}_\mu\boldsymbol{\mathfrak{p}}_\nu - \boldsymbol{\mathfrak{x}}_\mu\boldsymbol{\mathfrak{x}}_\nu) \\
\boldsymbol{\beth}_{\boldsymbol{\mu\nu}}^{\times} = \dfrac{1}{4}(\boldsymbol{\mathfrak{p}}_\mu\boldsymbol{\mathfrak{x}}_\nu + \boldsymbol{\mathfrak{x}}_\nu\boldsymbol{\mathfrak{p}}_\mu)
\end{cases}
\quad (64)
$$

If we define also the operators

$$
\begin{cases}
\boldsymbol{\mathfrak{z}}_\mu^{-} = \dfrac{1}{\sqrt{2}}(\boldsymbol{\mathfrak{p}}_\mu - i\boldsymbol{\mathfrak{x}}_\mu) \\
\boldsymbol{\mathfrak{z}}_\mu^{+} = \dfrac{1}{\sqrt{2}}(\boldsymbol{\mathfrak{p}}_\mu + i\boldsymbol{\mathfrak{x}}_\mu)
\end{cases}
\quad (65)
$$

We have

$$
\begin{cases}
\boldsymbol{\beth}_{\boldsymbol{\mu\nu}}^{+} = \dfrac{1}{4}(\boldsymbol{\mathfrak{z}}_\mu^{+}\boldsymbol{\mathfrak{z}}_\nu^{-} + \boldsymbol{\mathfrak{z}}_\mu^{-}\boldsymbol{\mathfrak{z}}_\nu^{+}) \\
\boldsymbol{\beth}_{\boldsymbol{\mu\nu}}^{-} = \dfrac{1}{4}(\boldsymbol{\mathfrak{z}}_\mu^{+}\boldsymbol{\mathfrak{z}}_\nu^{+} + \boldsymbol{\mathfrak{z}}_\mu^{-}\boldsymbol{\mathfrak{z}}_\nu^{-}) \\
\boldsymbol{\beth}_{\boldsymbol{\mu\nu}}^{\times} = \dfrac{i}{8}\left([\boldsymbol{\mathfrak{z}}_\mu^{+}, \boldsymbol{\mathfrak{z}}_\nu^{-}]_+ - [\boldsymbol{\mathfrak{z}}_\mu^{+}, \boldsymbol{\mathfrak{z}}_\nu^{+}]_+ + [\boldsymbol{\mathfrak{z}}_\mu^{-}, \boldsymbol{\mathfrak{z}}_\nu^{-}]_+ - [\boldsymbol{\mathfrak{z}}_\mu^{-}, \boldsymbol{\mathfrak{z}}_\nu^{+}]_+\right)
\end{cases}
\quad (66)
$$

$[\boldsymbol{\mathfrak{z}}_\mu^{+}, \boldsymbol{\mathfrak{z}}_\nu^{-}]_+$being the anticomutator

$$
[\boldsymbol{\mathfrak{z}}_\mu^{+}, \boldsymbol{\mathfrak{z}_\nu}^{-}]_+ = \boldsymbol{\mathfrak{z}}_\mu^{+}\boldsymbol{\mathfrak{z}}_\nu^{-} + \boldsymbol{\mathfrak{z}}_\nu^{-}\boldsymbol{\mathfrak{z}_\mu}^{+}
$$

From the commutation relation $[\boldsymbol{p_\mu}, \boldsymbol{x^\nu}]_- = i\delta_{\boldsymbol{\mu}}^{\boldsymbol{\nu}}$ we may deduce the following commutators

$$[\boldsymbol{p}_\mu, \boldsymbol{x}_\nu]_- = i\eta_{\mu\nu} \tag{67}$$

$$[\boldsymbol{z}_\mu^+, \boldsymbol{z}_\nu^-]_- = \eta_{\mu\nu} \tag{68}$$

$$\begin{cases} [\boldsymbol{x}_\mu \boldsymbol{x}_\nu, \boldsymbol{p}_\rho]_- = -i(\eta_{\nu\rho}\boldsymbol{x}_\mu - \eta_{\mu\rho}\boldsymbol{x}_\nu) \\ [\boldsymbol{p}_\mu \boldsymbol{x}_\nu, \boldsymbol{p}_\rho]_- = -i\eta_{\nu\rho}\boldsymbol{p}_\mu \\ [\boldsymbol{x}_\mu \boldsymbol{p}_\nu, \boldsymbol{p}_\rho]_- = -i\eta_{\mu\rho}\boldsymbol{p}_\nu \end{cases} \tag{69}$$

$$\begin{cases} [\boldsymbol{p}_\mu \boldsymbol{p}_\nu, \boldsymbol{x}_\rho]_- = i(\eta_{\nu\rho}\boldsymbol{p}_\mu + \eta_{\mu\rho}\boldsymbol{p}_\nu) \\ [\boldsymbol{p}_\mu \boldsymbol{x}_\nu, \boldsymbol{x}_\rho]_- = i\eta_{\mu\rho}\boldsymbol{x}_\nu \\ [\boldsymbol{x}_\mu \boldsymbol{p}_\nu, \boldsymbol{x}_\rho]_- = i\eta_{\nu\rho}\boldsymbol{x}_\mu \end{cases} \tag{70}$$

$$\begin{cases} [\daleth_{\mu\nu}^+, \boldsymbol{p}_\rho]_- = -\frac{i}{4}(\eta_{\nu\rho}\boldsymbol{x}_\mu + \eta_{\mu\rho}\boldsymbol{x}_\nu) \\ [\daleth_{\mu\nu}^-, \boldsymbol{p}_\rho]_- = \frac{i}{4}(\eta_{\nu\rho}\boldsymbol{x}_\mu + \eta_{\mu\rho}\boldsymbol{x}_\nu) \\ [\daleth_{\mu\nu}^\times, \boldsymbol{p}_\rho]_- = -\frac{i}{2}\eta_{\nu\rho}\boldsymbol{p}_\mu \end{cases} \tag{71}$$

$$\begin{cases} [\daleth_{\mu\nu}^+, \boldsymbol{x}_\rho]_- = \frac{i}{4}(\eta_{\nu\rho}\boldsymbol{p}_\mu + \eta_{\mu\rho}\boldsymbol{p}_\nu) \\ [\daleth_{\mu\nu}^-, \boldsymbol{x}_\rho]_- = \frac{i}{4}(\eta_{\nu\rho}\boldsymbol{p}_\mu + \eta_{\mu\rho}\boldsymbol{p}_\nu) \\ [\daleth_{\mu\nu}^\times, \boldsymbol{x}_\rho]_- = \frac{i}{2}\eta_{\mu\rho}\boldsymbol{x}_\nu \end{cases} \tag{72}$$

$$\begin{cases} [\boldsymbol{p}_\mu \boldsymbol{p}_\nu, \boldsymbol{x}_\rho \boldsymbol{x}_\lambda]_- = i(\eta_{\lambda\nu}\boldsymbol{p}_\mu\boldsymbol{x}_\rho + \eta_{\rho\nu}\boldsymbol{p}_\mu\boldsymbol{x}_\lambda + \eta_{\lambda\mu}\boldsymbol{x}_\rho\boldsymbol{p}_\nu + \eta_{\rho\mu}\boldsymbol{x}_\lambda\boldsymbol{p}_\nu) \\ [\boldsymbol{p}_\mu \boldsymbol{p}_\nu, \boldsymbol{p}_\rho \boldsymbol{x}_\lambda]_- = i(\eta_{\lambda\nu}\boldsymbol{p}_\mu\boldsymbol{p}_\rho + \eta_{\lambda\mu}\boldsymbol{p}_\rho\boldsymbol{p}_\nu) \\ [\boldsymbol{p}_\mu \boldsymbol{p}_\nu, \boldsymbol{x}_\rho \boldsymbol{p}_\lambda]_- = i(\eta_{\rho\nu}\boldsymbol{p}_\mu\boldsymbol{p}_\lambda + \eta_{\rho\mu}\boldsymbol{p}_\lambda\boldsymbol{p}_\nu) \\ [\boldsymbol{x}_\mu \boldsymbol{x}_\nu, \boldsymbol{p}_\rho \boldsymbol{x}_\lambda]_- = -i(\eta_{\rho\nu}\boldsymbol{x}_\mu\boldsymbol{x}_\lambda - \eta_{\rho\mu}\boldsymbol{x}_\lambda\boldsymbol{x}_\nu) \\ [\boldsymbol{x}_\mu \boldsymbol{x}_\nu, \boldsymbol{x}_\rho \boldsymbol{p}_\lambda]_- = -i(\eta_{\lambda\nu}\boldsymbol{x}_\mu\boldsymbol{x}_\rho - \eta_{\lambda\mu}\boldsymbol{x}_\rho\boldsymbol{x}_\nu) \\ [\boldsymbol{p}_\mu \boldsymbol{x}_\nu, \boldsymbol{p}_\rho \boldsymbol{x}_\lambda]_- = -i(\eta_{\rho\nu}\boldsymbol{p}_\mu\boldsymbol{x}_\lambda + \eta_{\lambda\mu}\boldsymbol{p}_\rho\boldsymbol{x}_\nu) \\ [\boldsymbol{p}_\mu \boldsymbol{x}_\nu, \boldsymbol{x}_\rho \boldsymbol{p}_\lambda]_- = -i(\eta_{\lambda\nu}\boldsymbol{p}_\mu\boldsymbol{x}_\rho + \eta_{\rho\mu}\boldsymbol{p}_\lambda\boldsymbol{x}_\nu) \\ [\boldsymbol{x}_\mu \boldsymbol{p}_\nu, \boldsymbol{x}_\rho \boldsymbol{p}_\lambda]_- = i(\eta_{\rho\nu}\boldsymbol{x}_\mu\boldsymbol{p}_\lambda - \eta_{\lambda\mu}\boldsymbol{x}_\rho\boldsymbol{p}_\nu) \end{cases} \tag{73}$$

$$\begin{cases} [\daleth_{\mu\nu}^+, \daleth_{\rho\lambda}^+]_- = \frac{i}{8}[\eta_{\nu\lambda}(\daleth_{\mu\rho}^\times - \daleth_{\rho\mu}^\times) + \eta_{\nu\rho}(\daleth_{\mu\lambda}^\times - \daleth_{\lambda\mu}^\times) + \eta_{\mu\lambda}(\daleth_{\nu\rho}^\times - \daleth_{\rho\nu}^\times) + \eta_{\mu\rho}(\daleth_{\nu\lambda}^\times - \daleth_{\lambda\nu}^\times)] \\ [\daleth_{\mu\nu}^-, \daleth_{\rho\lambda}^-]_- = -\frac{i}{8}[\eta_{\nu\lambda}(\daleth_{\mu\rho}^\times - \daleth_{\rho\mu}^\times) + \eta_{\nu\rho}(\daleth_{\mu\lambda}^\times - \daleth_{\lambda\mu}^\times) + \eta_{\mu\lambda}(\daleth_{\nu\rho}^\times - \daleth_{\rho\nu}^\times) + \eta_{\mu\rho}(\daleth_{\nu\lambda}^\times - \daleth_{\lambda\nu}^\times)] \\ [\daleth_{\mu\nu}^\times, \daleth_{\rho\lambda}^\times]_- = \frac{i}{2}(\eta_{\nu\mu}\daleth_{\rho\lambda}^\times - \eta_{\rho\lambda}\daleth_{\mu\nu}^\times) \end{cases} \tag{74}$$

$$\begin{cases} [\daleth_{\mu\nu}^+, \daleth_{\rho\lambda}^-]_- = \frac{i}{8}[\eta_{\nu\lambda}(\daleth_{\mu\rho}^\times + \daleth_{\rho\mu}^\times) + \eta_{\nu\rho}(\daleth_{\mu\lambda}^\times + \daleth_{\lambda\mu}^\times) + \eta_{\mu\lambda}(\daleth_{\nu\rho}^\times + \daleth_{\rho\nu}^\times) + \eta_{\mu\rho}(\daleth_{\nu\lambda}^\times + \daleth_{\lambda\nu}^\times)] \\ [\daleth_{\mu\nu}^-, \daleth_{\rho\lambda}^\times]_- = \frac{i}{4}[\eta_{\lambda\nu}(\daleth_{\mu\rho}^+ + \daleth_{\mu\rho}^-) + \eta_{\lambda\mu}(\daleth_{\rho\nu}^+ + \daleth_{\rho\nu}^-) + \eta_{\rho\nu}(\daleth_{\mu\lambda}^+ - \daleth_{\mu\lambda}^-) + \eta_{\rho\mu}(\daleth_{\lambda\nu}^+ - \daleth_{\lambda\nu}^-)] \\ [\daleth_{\mu\nu}^\times, \daleth_{\rho\lambda}^+]_- = -\frac{i}{4}[\eta_{\nu\lambda}(\daleth_{\mu\rho}^+ + \daleth_{\mu\rho}^-) + \eta_{\nu\rho}(\daleth_{\mu\lambda}^+ + \daleth_{\mu\lambda}^-) - \eta_{\mu\lambda}(\daleth_{\rho\nu}^+ - \daleth_{\rho\nu}^-) - \eta_{\mu\rho}(\daleth_{\nu\lambda}^+ - \daleth_{\nu\lambda}^-)] \end{cases} \tag{75}$$

It can be deduced easily from the relations (74) and (75) that the set $\{\daleth_{\mu\nu}^+, \daleth_{\mu\nu}^-, \daleth_{\mu\nu}^\times\}$ is a basis of a Lie algebra which is the dispersion operators algebra for the multidimensional case.
As the indices $\mu, \nu, \rho, \lambda$ run from 0 to $N-1$, the dimension $D$ of this dispersion operators algebra which is equal to the numbers of the elements of the basis $\{\daleth_{\mu\nu}^+, \daleth_{\mu\nu}^-, \daleth_{\mu\nu}^\times\}$ is

$$D = \frac{N(N+1)}{2} + \frac{N(N+1)}{2} + N^2 = N(2N+1) \tag{76}$$

In fact, from the relation (64) we can deduce that the number of operators $\daleth_{\mu\nu}^+$is equal to $\frac{N(N+1)}{2}$, the number of operators $\daleth_{\mu\nu}^-$ equal to $\frac{N(N+1)}{2}$ and the number of operators $\daleth_{\mu\nu}^\times$ equal to $N^2$.

### 4.2 Linear Canonical Transformations

We may define the linear canonical transformation as the linear transformation given by the relation

$$\begin{cases} \boldsymbol{p}_\mu{}' = \Pi_\mu^\nu \boldsymbol{p}_\nu + \Theta_\mu^\nu \boldsymbol{x}_\nu \\ \boldsymbol{x}_\mu{}' = \Xi_\mu^\nu \boldsymbol{p}_\nu + \Lambda_\mu^\nu \boldsymbol{x}_\nu \end{cases} \Leftrightarrow (\boldsymbol{p}' \quad \boldsymbol{x}') = (\boldsymbol{p} \quad \boldsymbol{x}) \begin{pmatrix} \Pi & \Xi \\ \Theta & \Lambda \end{pmatrix} \tag{77}$$

and which leave invariant the canonical commutation relations

$$\begin{cases} \left[\boldsymbol{p}'_\mu, \boldsymbol{p}'_\nu\right]_- = \left[\boldsymbol{p}_\mu, \boldsymbol{p}_\nu\right]_- = 0 \\ \left[\boldsymbol{x}'_\mu, \boldsymbol{x}'_\nu\right]_- = \left[\boldsymbol{x}_\mu, \boldsymbol{x}_\nu\right]_- = 0 \\ \left[\boldsymbol{p}'_\mu, \boldsymbol{x}'_\nu\right]_- = \left[\boldsymbol{p}_\mu, \boldsymbol{x}_\nu\right]_- = i\eta_{\mu\nu} \end{cases} \tag{78}$$

we obtain the following conditions

$$\begin{cases} \Pi_\mu^\rho \eta_{\rho\lambda} \Theta_\nu^\lambda - \Theta_\mu^\rho \eta_{\rho\lambda} \Pi_\nu^\lambda = 0 \\ \Xi_\mu^\rho \eta_{\rho\lambda} \Lambda_\nu^\lambda - \Lambda_\mu^\rho \eta_{\rho\lambda} \Xi_\nu^\lambda = 0 \\ \Pi_\mu^\rho \eta_{\rho\lambda} \Lambda_\nu^\lambda - \Xi_\nu^\lambda \eta_{\rho\lambda} \Theta_\mu^\rho = \eta_{\mu\nu} \end{cases} \Leftrightarrow \begin{cases} \Pi^t \eta \Theta - \Theta^t \eta \Pi = 0 \\ \Xi^t \eta \Lambda - \Lambda^t \eta \Xi = 0 \\ \Pi^t \eta \Lambda - \Theta^t \eta \Xi = \eta \end{cases} \Leftrightarrow \begin{pmatrix} \Pi & \Xi \\ \Theta & \Lambda \end{pmatrix}^t \begin{pmatrix} 0 & \eta \\ -\eta & 0 \end{pmatrix} \begin{pmatrix} \Pi & \Xi \\ \Theta & \Lambda \end{pmatrix} = \begin{pmatrix} 0 & \eta \\ -\eta & 0 \end{pmatrix} \tag{79}$$

If the signature of $\eta$ is $(N, 0)$ , it is equal to the $N \times N$ identity matrix $\eta = I_N$ (case of Euclidian space), the relation (79) become

$$\begin{pmatrix} \Pi & \Xi \\ \Theta & \Lambda \end{pmatrix}^t \begin{pmatrix} 0 & I_N \\ -I_N & 0 \end{pmatrix} \begin{pmatrix} \Pi & \Xi \\ \Theta & \Lambda \end{pmatrix} = \begin{pmatrix} 0 & I_N \\ -I_N & 0 \end{pmatrix} \tag{80}$$

according to this relation the $2N \times 2N$ matrix $\begin{pmatrix} \Pi & \Xi \\ \Theta & \Lambda \end{pmatrix}$ is in this case a symplectic matrix i.e an element of the symplectic group $Sp\ (2N)$. We may generalize this result for the general case of pseudo-Euclidian space i.e with $\eta$ having a signature $(N_+, N_-)$ : in that case, we may call a matrix $\begin{pmatrix} \Pi & \Xi \\ \Theta & \Lambda \end{pmatrix}$ verifying the relation (79) a pseudo-symplectic matrix and their set the pseudo-symplectic group. We may denote this Lie group $Sp\ (2N_+, 2N_-)$ and its Lie algebra $\mathfrak{sp}(2N_+, 2N_-)$. The matrix $\begin{pmatrix} \Pi & \Xi \\ \Theta & \Lambda \end{pmatrix}$ can be written in the form

$$\begin{pmatrix} \Pi & \Xi \\ \Theta & \Lambda \end{pmatrix} = e^{\mathcal{M}} = e^{\begin{pmatrix} \mathcal{M}_1 & \mathcal{M}_3 \\ \mathcal{M}_2 & \mathcal{M}_4 \end{pmatrix}} \tag{81}$$

in which $\mathcal{M} = \begin{pmatrix} \mathcal{M}_1 & \mathcal{M}_3 \\ \mathcal{M}_2 & \mathcal{M}_4 \end{pmatrix}$ is an element of the Lie algebra $\mathfrak{sp}(2N_+, 2N_-)$, we have

$$\begin{pmatrix} \Pi & \Xi \\ \Theta & \Lambda \end{pmatrix}^t \begin{pmatrix} 0 & \eta \\ -\eta & 0 \end{pmatrix} \begin{pmatrix} \Pi & \Xi \\ \Theta & \Lambda \end{pmatrix} = \begin{pmatrix} 0 & \eta \\ -\eta & 0 \end{pmatrix} \Leftrightarrow \begin{cases} \mathcal{M}_2^t = \eta \mathcal{M}_2 \eta \\ \mathcal{M}_3^t = \eta \mathcal{M}_3 \eta \\ \mathcal{M}_4 = -\eta \mathcal{M}_1^t \eta \end{cases} \tag{82}$$

It can be deduced easily from the relation (82) that the matrix $\mathcal{M}$ and his transpose $\mathcal{M}^t$ are of the form

$$\mathcal{M} = \begin{pmatrix} \mathcal{M}_1 & \mathcal{M}_3 \\ \mathcal{M}_2 & -\eta \mathcal{M}_1^t \eta \end{pmatrix} \qquad \mathcal{M}^t = \begin{pmatrix} \mathcal{M}_1^t & \eta \mathcal{M}_2 \eta \\ \eta \mathcal{M}_3 \eta & -\eta \mathcal{M}_1 \eta \end{pmatrix} \tag{83}$$

If we introduce the parametrization

$$\begin{cases} \mathcal{M}_1 = \eta \mathcal{X} \\ \mathcal{M}_2 = \eta \mathcal{Y} \\ \mathcal{M}_3 = \eta \mathcal{Z} \end{cases} \tag{84}$$

We obtain for $\mathcal{M}$ and $\mathcal{M}^t$

$$\mathcal{M} = \begin{pmatrix} \eta \mathcal{X} & \eta \mathcal{Z} \\ \eta \mathcal{Y} & -\eta \mathcal{X}^t \end{pmatrix} \qquad \mathcal{M}^t = \begin{pmatrix} \mathcal{X}^t \eta & \mathcal{Y} \eta \\ \mathcal{Z} \eta & -\mathcal{X} \eta \end{pmatrix} \tag{85}$$

Then for an infinitesimal linear canonical transformation, we have

$$(\boldsymbol{p}' \quad \boldsymbol{x}') = (\boldsymbol{p} \quad \boldsymbol{x})(1+\mathcal{M}) = (\boldsymbol{p} \quad \boldsymbol{x})\begin{pmatrix} 1+\eta\mathcal{X} & \eta\mathcal{Z} \\ \eta\mathcal{Y} & 1-\eta\mathcal{X}^t \end{pmatrix}$$

$$\begin{cases} \boldsymbol{p}' = \boldsymbol{p} + \eta\mathcal{X}\boldsymbol{p} + \eta\mathcal{Y}\boldsymbol{x} \\ \boldsymbol{x}' = \boldsymbol{x} + \eta\mathcal{Z}\boldsymbol{p} - \eta\mathcal{X}^t\boldsymbol{x} \end{cases} \Leftrightarrow \begin{cases} \boldsymbol{p}'_{\boldsymbol{\mu}} = \boldsymbol{p}_{\boldsymbol{\mu}} + [\eta\mathcal{X}]_\mu^\nu \boldsymbol{p}_{\boldsymbol{\nu}} + [\eta\mathcal{Y}]_\mu^\nu \boldsymbol{x}_{\boldsymbol{\nu}} \\ \boldsymbol{x}'_{\boldsymbol{\mu}} = \boldsymbol{x}_{\boldsymbol{\mu}} + [\eta\mathcal{Z}]_\mu^\nu \boldsymbol{p}_{\boldsymbol{\nu}} - [\eta\mathcal{X}^t]_\nu^\mu \boldsymbol{x}_{\boldsymbol{\nu}} \end{cases}$$

$$\begin{cases} \boldsymbol{p}'_{\boldsymbol{\mu}} = \boldsymbol{p}_{\boldsymbol{\mu}} + \eta_{\mu\rho}\mathcal{X}^{\rho\nu}\boldsymbol{p}_{\boldsymbol{\nu}} + \eta_{\mu\rho}\mathcal{Y}^{\rho\nu}\boldsymbol{x}_\nu \\ \boldsymbol{x}'_{\boldsymbol{\mu}} = \boldsymbol{x}_{\boldsymbol{\mu}} + \eta_{\mu\rho}\mathcal{Z}^{\rho\nu}\boldsymbol{p}_{\boldsymbol{\nu}} - \eta_{\mu\rho}\mathcal{X}^{\nu\rho}\boldsymbol{x}_{\boldsymbol{\nu}} \end{cases} \quad (86)$$

We may introduce a unitary representation of the linear canonical transformation

$$\begin{cases} \boldsymbol{p}'_{\boldsymbol{\mu}} = \boldsymbol{U}\boldsymbol{p}_{\boldsymbol{\mu}}\boldsymbol{U}^\dagger \\ \boldsymbol{x}'_{\boldsymbol{\mu}} = \boldsymbol{U}\boldsymbol{x}_{\boldsymbol{\mu}}\boldsymbol{U}^\dagger \end{cases} \quad (87)$$

and we can verify that $\boldsymbol{U}$ and$\boldsymbol{U}^\dagger$can be given by the relations

$$\begin{cases} \boldsymbol{U} = e^{i(\theta_+^{\rho\lambda}\daleth_{\boldsymbol{\rho\lambda}}^+ + \theta_-^{\rho\lambda}\daleth_{\boldsymbol{\rho\lambda}}^- + \theta_\times^{\rho\lambda}\daleth_{\boldsymbol{\rho\lambda}}^\times)} \\ \boldsymbol{U}^\dagger = e^{-i(\theta_+^{\epsilon\sigma}\daleth_{\epsilon\sigma}^+ + \theta_-^{\epsilon\sigma}\daleth_{\epsilon\sigma}^- + \theta_\times^{\epsilon\sigma}\daleth_{\epsilon\sigma}^\times)} \end{cases} \quad (88)$$

In fact, for an infinitesimal transformation

$$\begin{cases} \boldsymbol{U} = 1 + i(\theta_+^{\rho\lambda}\daleth_{\boldsymbol{\rho\lambda}}^+ + \theta_-^{\rho\lambda}\daleth_{\boldsymbol{\rho\lambda}}^- + \theta_\times^{\rho\lambda}\daleth_{\boldsymbol{\rho\lambda}}^\times) \\ \boldsymbol{U}^\dagger = 1 - i(\theta_+^{\epsilon\sigma}\daleth_{\epsilon\sigma}^+ + \theta_-^{\epsilon\sigma}\daleth_{\epsilon\sigma}^- + \theta_\times^{\epsilon\sigma}\daleth_{\epsilon\sigma}^\times) \end{cases} \quad (89)$$

So (87) become

$$\begin{cases} \boldsymbol{p}'_{\boldsymbol{\mu}} = \boldsymbol{p}_{\boldsymbol{\mu}} + i\theta_+^{\rho\lambda}[\daleth_{\boldsymbol{\rho\lambda}}^+, \boldsymbol{p}_{\boldsymbol{\mu}}]_- + i\theta_-^{\rho\lambda}[\daleth_{\boldsymbol{\rho\lambda}}^-, \boldsymbol{p}_{\boldsymbol{\mu}}]_- + i\theta_\times^{\rho\lambda}[\daleth_{\boldsymbol{\rho\lambda}}^\times, \boldsymbol{p}_{\boldsymbol{\mu}}]_- \\ \boldsymbol{x}'_{\boldsymbol{\mu}} = \boldsymbol{x}_{\boldsymbol{\mu}} + i\theta_+^{\rho\lambda}[\daleth_{\boldsymbol{\rho\lambda}}^+, \boldsymbol{x}_{\boldsymbol{\mu}}]_- + i\theta_-^{\rho\lambda}[\daleth_{\boldsymbol{\rho\lambda}}^-, \boldsymbol{x}_{\boldsymbol{\mu}}]_- + i\theta_\times^{\rho\lambda}[\daleth_{\boldsymbol{\rho\lambda}}^\times, \boldsymbol{x}_{\boldsymbol{\mu}}]_- \end{cases} \quad (90)$$

Then, taking into account the relations (71) and (72), we obtain

$$\begin{cases} \boldsymbol{p}'_{\boldsymbol{\mu}} = \boldsymbol{p}_{\boldsymbol{\mu}} + \frac{1}{2}\eta_{\rho\mu}(\theta_+^{\rho\nu} - \theta_-^{\rho\nu})\boldsymbol{x}_\nu + \frac{1}{2}\theta_\times^{\nu\rho}\eta_{\rho\mu}\boldsymbol{p}_{\boldsymbol{\nu}} \\ \boldsymbol{x}'_{\boldsymbol{\mu}} = \boldsymbol{x}_{\boldsymbol{\mu}} - \frac{1}{2}\eta_{\rho\mu}(\theta_+^{\rho\nu} + \theta_-^{\nu\rho})\boldsymbol{p}_\nu - \frac{1}{2}\eta_{\rho\mu}\theta_\times^{\rho\nu}\boldsymbol{x}_{\boldsymbol{\nu}} \end{cases} \quad (91)$$

Indefying the relations (86) and (91) gives

$$\begin{cases} \mathcal{X}^{\rho\nu} = \frac{1}{2}\theta_\times^{\nu\rho} \\ \mathcal{Y}^{\rho\nu} = \frac{1}{2}\left(\theta_+^{\rho\nu} - \theta_-^{\rho\nu}\right) \\ \mathcal{Z}^{\rho\nu} = -\frac{1}{2}\left(\theta_+^{\rho\nu} + \theta_-^{\rho\nu}\right) \end{cases} \Leftrightarrow \begin{cases} \mathcal{X} = \frac{1}{2}\theta_\times^t \\ \mathcal{Y} = \frac{1}{2}(\theta_+ - \theta_-) \\ \mathcal{Z} = -\frac{1}{2}(\theta_+ - \theta_-) \end{cases} \quad (92)$$

then the relations in (83) and (85) become

$$\mathcal{M} = \begin{pmatrix} \mathcal{M}_1 & \mathcal{M}_3 \\ \mathcal{M}_2 & -\eta\mathcal{M}_1^t\eta \end{pmatrix} = \begin{pmatrix} \eta\mathcal{X} & \eta\mathcal{Z} \\ \eta\mathcal{Y} & -\eta\mathcal{X}^t \end{pmatrix} = \frac{1}{2}\begin{pmatrix} \eta\theta_\times^t & -\eta(\theta_+ + \theta_-) \\ \eta(\theta_+ - \theta_-) & -\eta\theta_\times \end{pmatrix} \quad (93)$$

$$\mathcal{M}^t = \begin{pmatrix} \mathcal{M}_1^t & \eta\mathcal{M}_2\eta \\ \eta\mathcal{M}_3\eta & -\eta\mathcal{M}_1\eta \end{pmatrix} = \begin{pmatrix} \mathcal{X}^t\eta & \mathcal{Y}\eta \\ \mathcal{Z}\eta & -\mathcal{X}\eta \end{pmatrix} = \frac{1}{2}\begin{pmatrix} \theta_\times\eta & (\theta_+ - \theta_-)\eta \\ -(\theta_+ - \theta_-)\eta & -\theta_\times^t\eta \end{pmatrix} \quad (94)$$

So briefly, we have for a linear canonical transformation in the case of multidimensional theory

$$\begin{cases} \boldsymbol{p}_{\boldsymbol{\mu}}' = \Pi_\mu^\nu \boldsymbol{p}_{\boldsymbol{\nu}} + \Theta_\mu^\nu \boldsymbol{x}_{\boldsymbol{\nu}} = \boldsymbol{U}\boldsymbol{p}_{\boldsymbol{\mu}}\boldsymbol{U}^\dagger \\ \boldsymbol{x}_{\boldsymbol{\mu}}' = \Xi_\mu^\nu \boldsymbol{p}_{\boldsymbol{\nu}} + \Lambda_\mu^\nu \boldsymbol{x}_{\boldsymbol{\nu}} = \boldsymbol{U}\boldsymbol{x}_{\boldsymbol{\mu}}\boldsymbol{U}^\dagger \end{cases} \Leftrightarrow (\boldsymbol{p}' \quad \boldsymbol{x}') = (\boldsymbol{p} \quad \boldsymbol{x})\begin{pmatrix} \Pi & \Xi \\ \Theta & \Lambda \end{pmatrix} \quad (95)$$

with

$$\begin{pmatrix} \Pi & \Xi \\ \Theta & \Lambda \end{pmatrix} = e^{\mathcal{M}} = e^{\frac{1}{2}\begin{pmatrix} \eta\theta_\times^t & -\eta(\theta_+ + \theta_-) \\ \eta(\theta_+ - \theta_-) & -\eta\theta_\times \end{pmatrix}} \quad (96)$$

and

$$\boldsymbol{U} = e^{i(\theta_+^{\rho\lambda}\daleth_{\boldsymbol{\rho\lambda}}^+ + \theta_-^{\rho\lambda}\daleth_{\boldsymbol{\rho\lambda}}^- + \theta_\times^{\rho\lambda}\daleth_{\boldsymbol{\rho\lambda}}^\times)} \quad (97)$$

## 5 Conclusions

The results obtained in the sections 2 and 3 show that the introduction of the dispersion operators algebra permits to perform a natural and well description of the link which can be established between the phase space representation of quantum mechanics and linear canonical transformation. This link is a consequence of the existence of relationship between dispersion operators and the phase space representation on one hand and dispersion operator algebra and linear canonical transformation on the other hand. The phase space representation is built with the eigeinstates $|n, X, P, \mathcal{b}\rangle$ of dispersion operators; linear canonical transformation can be represented using the dispersion operators algebra. The relations (32), (33) and (53) allows to conclude that a right way to describe and to represent linear canonical transformation over state space in framework of quantum mechanics is to use the basis $\{|n, X, P, \mathcal{b}\rangle\}$.

The calculations performed in the section 4 show that these main results obtained for the case of one dimension quantum mechanics may be generalized in the case of multidimensional theory.

The results that we have established in this paper may have many interesting applications in various scientific and technical fields.

## References

1. Ravo Tokiniaina Ranaivoson : Raoelina Andriambololona, Rakotoson Hanitriarivo, Roland Raboanary: Study on a Phase Space Representation of Quantum Theory, arXiv:1304.1034 [quant-ph], International Journal of Latest Research in Science and Technology, ISSN(Online):2278-5299, Volume 2,Issue 2 :Page No.26-35,March-April, 2013
2. Raoelina Andriambololona, Ravo Tokiniaina Ranaivoson, Rakotoson Hanitriarivo, Wilfrid Chrysante Solofoarisina : Study on Linear Canonical Transformation in a Framework of a Phase Space Representation of Quantum Mechanics, arXiv: 1503.02449 [quant-ph], International Journal of Applied Mathematics and Theoretical Physics. Vol. 1, No. 1, 2015, pp. 1-8, 2015
3. E.P. Wigner : On the quantum correction for thermodynamic equilibrium, Phys. Rev 40, 749-759, 1932
4. H.J. Groenewold : On the Principles of elementary quantum mechanics, Physica, Volume 12, Issue 7, 1946
5. J.E. Moyal : Quantum mechanics as a statistical theory, Proceedings of the Cambridge Philosophical Society 45, 99–124, 1949
6. G Torres-Vega, J.H. Frederick : A quantum mechanical representation in phase space , J. Chem. Phys. 98 (4), 1993
7. H.-W. Lee: Theory and application of the quantum phase-space distribution functions, Phys.Rep 259, Issue 3, 147-211, 1995
8. K. B Møller, T. G Jørgensen , G. Torres-Vega : On coherent-state representations of quantum mechanics: Wave mechanics in phase space. Journal of Chemical Physics, 106(17), 7228-7240. DOI: 10.1063/1.473684, 1997
9. A. Nassimi : Quantum Mechanics in Phase Space, arXiv:0706.0237 [quant-ph], 2008
10. T.L Curtright ,C.K. Zachos : Quantum Mechanics in Phase Space, arXiv:1104.5269 [physics.hist-ph], Asia Pacific Physics Newsletter, V1, Iss 1, pp 37-46, May 2012
11. D. K. Ferry : Phase-space functions: can they give a different view of quantum Mechanics, Journal of Computational Electronics, Volume 14, Issue 4, pp 864-868, December 2015.
12. M. Moshinsky and C. Quesne : Linear canonical transformations and their unitary representations, J. Math. Phys.12, 8, 1772–1783, 1971
13. Arlen Anderson : Canonical Transformations in Quantum Mechanics, Annals Phys. 232 (1994) 292-331
14. Tian-Zhou Xu, Bing-Zhao Li: Linear Canonical Transform and Its Applications, Science Press, Beijing, China, 2013.
15. Raoelina Andriambololona : Algèbre linéaire et multilinéaire, Applications, Collection LIRA, INSTN Madagascar,1986
16. Ravo Tokiniaina Ranaivoson, Raoelina Andriambololona, Rakotoson Hanitriarivo. : Time- Frequency analysis and harmonic Gaussian functions, Pure and Applied Mathematics Journal.Vol. 2, No. 2,2013, pp. 71-78. doi: 10.11648/j.pamj.20130202.14
17. Raoelina Andriambololona : Mecanique quantique, Collection LIRA, INSTN Madagascar, 1990
18. K . B. Wolf : A Top-Down Account of Linear Canonical Transforms , arXiv:1206.1123 [Math.ph], SIGMA 8 (2012), 033, 13 pages